\documentclass[aps,preprint, floatfix]{revtex4}
\usepackage{graphicx}
\usepackage[scanall]{psfrag}
\usepackage{amssymb}

\begin{document}

%		DEFINITIONS FOR TEX
%
%%%%%%%%%%%%%%%%%%%%%%%%%%%%%%%%%%%%%%%%%%%%%%%%%%%%%%%%%%%%%%%
%
%
%\def\e{\'e}
%\def\ee{\`e}
%%%%%%%%%%%%%%%%%%%DEFINITIONS%%%%%%%%%%%%%%%%%%%%%%%%%%%%%%%%%
%
\def\oti{{\otimes}}
\def\lb{ \left[ }
\def\rb{ \right]  }
\def\tilde{\widetilde}
\def\bar{\overline}
\def\hat{\widehat}
\def\*{\star}
\def\[{\left[}
\def\]{\right]}
\def\({\left(}		\def\BL{\Bigr(}
\def\){\right)}		\def\BR{\Bigr)}
	\def\BBL{\lb}
	\def\BBR{\rb}
%
%%%%%%%%%%%%%%%%%%%%%%%%%%%%%%%%%%%%%%%%%%%%%%%%%%%%%%%%%%%%%%%
%
\def\zb{{\bar{z} }}
\def\zbar{{\bar{z} }}
\def\frac#1#2{{#1 \over #2}}
\def\inv#1{{1 \over #1}}
\def\half{{1 \over 2}}
\def\d{\partial}
\def\der#1{{\partial \over \partial #1}}
\def\dd#1#2{{\partial #1 \over \partial #2}}
\def\vev#1{\langle #1 \rangle}
\def\ket#1{ | #1 \rangle}
\def\rvac{\hbox{$\vert 0\rangle$}}
\def\lvac{\hbox{$\langle 0 \vert $}}
\def\2pi{\hbox{$2\pi i$}}
\def\e#1{{\rm e}^{^{\textstyle #1}}}
\def\grad#1{\,\nabla\!_{{#1}}\,}
\def\dsl{\raise.15ex\hbox{/}\kern-.57em\partial}
\def\Dsl{\,\raise.15ex\hbox{/}\mkern-.13.5mu D}
%
%%%%%%%%%%%%%%%%%%%%GREEK LETTERS%%%%%%%%%%%%%%%%%%%%%%%%%%%%%%
%
%\def\th{\theta}		\def\Th{\Theta}
\def\ga{\gamma}		\def\Ga{\Gamma}
\def\be{\beta}
\def\al{\alpha}
\def\ep{\epsilon}
\def\vep{\varepsilon}
\def\la{\lambda}	\def\La{\Lambda}
\def\de{\delta}		\def\De{\Delta}
\def\om{\omega}		\def\Om{\Omega}
\def\sig{\sigma}	\def\Sig{\Sigma}
\def\vphi{\varphi}

%
%%%%%%%%%%%%%%%%%%%CALIGRAPHIC LETTERS%%%%%%%%%%%%%%%%%%%%%%%%%
%
\def\CA{{\cal A}}	\def\CB{{\cal B}}	\def\CC{{\cal C}}
\def\CD{{\cal D}}	\def\CE{{\cal E}}	\def\CF{{\cal F}}
\def\CG{{\cal G}}	\def\CH{{\cal H}}	\def\CI{{\cal J}}
\def\CJ{{\cal J}}	\def\CK{{\cal K}}	\def\CL{{\cal L}}
\def\CM{{\cal M}}	\def\CN{{\cal N}}	\def\CO{{\cal O}}
\def\CP{{\cal P}}	\def\CQ{{\cal Q}}	\def\CR{{\cal R}}
\def\CS{{\cal S}}	\def\CT{{\cal T}}	\def\CU{{\cal U}}
\def\CV{{\cal V}}	\def\CW{{\cal W}}	\def\CX{{\cal X}}
\def\CY{{\cal Y}}	\def\CZ{{\cal Z}}

\def\rvac{\hbox{$\vert 0\rangle$}}
\def\lvac{\hbox{$\langle 0 \vert $}}
\def\comm#1#2{ \BBL\ #1\ ,\ #2 \BBR }
\def\2pi{\hbox{$2\pi i$}}
\def\e#1{{\rm e}^{^{\textstyle #1}}}
\def\grad#1{\,\nabla\!_{{#1}}\,}
\def\dsl{\raise.15ex\hbox{/}\kern-.57em\partial}
\def\Dsl{\,\raise.15ex\hbox{/}\mkern-.13.5mu D}
%
%%%%%%%%%%%%%%%%%%%%GREEK LETTERS%%%%%%%%%%%%%%%%%%%%%%%%%%%%%%
%
%%%%%%%%%%%%%%% MATH CHARACTERS %%%%%%%%%%%%%%%%%%%%%%%%%%%%
%
\font\numbers=cmss12
%\font\numbers=cmu10 scaled\magstep1
\font\upright=cmu10 scaled\magstep1
\def\stroke{\vrule height8pt width0.4pt depth-0.1pt}
\def\topfleck{\vrule height8pt width0.5pt depth-5.9pt}
\def\botfleck{\vrule height2pt width0.5pt depth0.1pt}
\def\Zmath{\vcenter{\hbox{\numbers\rlap{\rlap{Z}\kern
0.8pt\topfleck}\kern 2.2pt
                   \rlap Z\kern 6pt\botfleck\kern 1pt}}}
\def\Qmath{\vcenter{\hbox{\upright\rlap{\rlap{Q}\kern
                   3.8pt\stroke}\phantom{Q}}}}
\def\Nmath{\vcenter{\hbox{\upright\rlap{I}\kern 1.7pt N}}}
\def\Cmath{\vcenter{\hbox{\upright\rlap{\rlap{C}\kern
                   3.8pt\stroke}\phantom{C}}}}
\def\Rmath{\vcenter{\hbox{\upright\rlap{I}\kern 1.7pt R}}}
\def\Z{\ifmmode\Zmath\else$\Zmath$\fi}
\def\Q{\ifmmode\Qmath\else$\Qmath$\fi}
\def\N{\ifmmode\Nmath\else$\Nmath$\fi}
\def\C{\ifmmode\Cmath\else$\Cmath$\fi}
\def\R{\ifmmode\Rmath\else$\Rmath$\fi}
%%%%%%%%%%%%%%%%%%%%%%%%%%%%%%%%%%%%%%%%%%%%%%%%%%%%%%%%%%%%%%%%%
 %%%%%%%%%%%%%%%%%% END OF DEFINITIONS %%%%%%%%%%%%%%%%%%%%%%
 %%%%%%%%%%%%%%%%%%%%%%%%%%%%%%%%%%%%%%%%%%%%%%%%%

\def\barray{\begin{eqnarray}}
\def\earray{\end{eqnarray}}
\def\beq{\begin{equation}}
\def\eeq{\end{equation}}

\def\n{\noindent}

\def\Tr{\rm Tr} 
\def\xvec{{\bf x}}
\def\kvec{{\bf k}}
\def\kvecp{{\bf k'}}
\def\omk{\om{\kvec}} 
\def\dk#1{\frac{d\kvec_{#1}}{(2\pi)^d}}
\def\2pid{(2\pi)^d}
\def\ket#1{|#1 \rangle}
\def\bra#1{\langle #1 |}
\def\vol{V}
\def\adag{a^\dagger}
\def\rme{{\rm e}}
\def\Im{{\rm Im}}
\def\pvec{{\bf p}}
\def\fermiS{\CS_F}
\def\cdag{c^\dagger}
\def\adag{a^\dagger}
\def\bdag{b^\dagger}
\def\vvec{{\bf v}}
\def\muhat{{\hat{\mu}}}
\def\vac{|0\rangle}
\def\pcut{{\Lambda_c}}
\def\chidot{\dot{\chi}}
\def\gradvec{\vec{\nabla}}
\def\psitilde{\tilde{\Psi}}
\def\psibar{\bar{\psi}}
\def\psidag{\psi^\dagger} 
\def\m{m_*}
\def\up{\uparrow}
\def\down{\downarrow}
\def\Qo{Q^{0}}
\def\vbar{\bar{v}}
\def\ubar{\bar{u}}
\def\smallhalf{{\textstyle \inv{2}}}
\def\smallsqrt{{\textstyle \inv{\sqrt{2}}}}
\def\rvec{{\bf r}}
\def\avec{{\bf a}}
\def\pivec{{\vec{\pi}}}
\def\svec{\vec{s}} 
\def\phivec{\vec{\phi}}
\def\daggerc{{\dagger_c}}
\def\Gfour{G^{(4)}}
\def\dim#1{\lbrack\!\lbrack #1 \rbrack\! \rbrack }
\def\qhat{{\hat{q}}}
\def\ghat{{\hat{g}}}
\def\nvec{{\vec{n}}}
\def\bull{$\bullet$}
\def\ghato{{\hat{g}_0}}
\def\r{r}
\def\deltaq{\delta_q}
\def\gcharge{g_q}
\def\gspin{g_s}
\def\deltas{\delta_s}
\def\gQC{g_{AF}} 
\def\ghatqc{\ghat_{AF}}
\def\xqc{x_{AF}}
\def\mhat{\hat{m}}
\def\xup{x_2}
\def\xdown{x_1}
\def\sigmavec{\vec{\sigma}}
\def\xopt{x_{\rm opt}}
\def\Lambdac{{\Lambda_c}}
\def\angstrom{{{\scriptstyle \circ} \atop A}     }
\def\AA{\leavevmode\setbox0=\hbox{h}\dimen0=\ht0 \advance\dimen0 by-1ex\rlap{
\raise.67\dimen0\hbox{\char'27}}A}
\def\ratio{\gamma}
\def\Phivec{{\vec{\Phi}}}
\def\singlet{\chi^- \chi^+} 
\def\mhat{{\hat{m}}}

\def\Im{{\rm Im}}

\def\xstar{x_*}

\title{Non-Fermi liquid properties of $2d$ symplectic fermions: 
the role of a dynamically generated (pseudo)-gap}
\author{Eliot Kapit  and Andr\'e  LeClair}
\affiliation{Newman Laboratory, Cornell University, Ithaca, NY}

\bigskip\bigskip\bigskip\bigskip

\begin{abstract}

The interacting
symplectic fermion model in two spatial dimensions is further
analyzed. 
 As an effective low energy theory,
the model is unitary.    
 We show  that a relativistic mass $m$  is  dynamically generated
and derive a gap equation for it.
By incorporating a  finite temperature we 
study some fundamental properties of the model, such 
as the specific heat and spin response,  which clearly
show non-Fermi liquid properties. We find that various physical properties
are suppressed at temperatures $T< T^*$ where
the cross-over scale is $T^* = m$.         As a simplified, toy  model
of high $T_c$ superconductivity,  
we thus  identify  the pseudogap energy scale with the zero temperature
relativistic mass $m$,  and show that this reproduces 
some qualitative aspects of 
the observed phenomenology of the pseudogap.  
The  effects of the pseudogap and finite temperature
on the d-wave gap equation are analyzed.   In this model, the pseudogap 
is a distinct phenomenon from superconductivity and in fact
competes with it.   
Our analysis of $T_c$ suggests that the quantum critical point of our
model, where the pseudogap vanishes,  occurs inside the superconducting dome
near optimal doping.   For an antiferromagnetic exchange energy of
$J/k_B \sim 1350K$, solutions of the d-wave gap equation give a
maximum $T_c$ of about $110K$.

\end{abstract}

\maketitle

\section{Introduction}

An important aspect of strongly-correlated electron physics 
is the expectation that some systems may exhibit
 novel non-Fermi liquid behavior.   This can present some interesting
challenges for theoretical models since non-Fermi liquid behavior 
is known to be rare based on simple renormalization group (RG)
 arguments\cite{Shankar}.  
The best understood example is the Luttinger liquid in one spatial dimension,
where the non-Fermi liquid behavior is essentially attributed to the
fact that quartic interactions of Dirac fermions are marginal in the RG sense. 
In higher dimensions quartic interactions of Dirac fermions are irrelevant.   
These considerations were one of the primary initial motivations for
the construction and analysis of a new  model of interacting 
fermions\cite{LeClair1,Neubert,KapitLeclair} with non-Fermi liquid behavior.
Like the Luttinger liquid, the model is a continuum
field theory of only 4 fermionic fields $\chi^\pm_{\up, \down}$ which
carry charge and spin, with a unique quartic interaction due to Fermi statistics. 
The novelty of the model is the free kinetic term in the hamiltonian 
which in addition the commonplace term that is second order in spatial 
derivatives $\gradvec \chi^- \cdot \gradvec \chi^+$,  contains an additional
contribution that is second order in time derivatives $\d_t \chi^- \d_t \chi^+$, 
and thus has an emergent Lorentz symmetry.   The model can be consistently
canonically quantized as a fermion, even though this structure of the kinetic
term is usually associated with relativistic bosons.   
  In two spatial dimensions the field
$\chi$ has dimension $1/2$, which implies the quartic interaction has dimension
$2$ and is thus RG relevant.    The model in fact has a low-energy fixed point,
i.e. quantum critical point.

Several interesting properties in addition to  the non-Fermi liquid behavior 
emerged in the analysis of the model\cite{KapitLeclair}. 
It  has a hidden $SO(5)$ symmetry 
which contains the spin $SO(3)$ and electric $U(1)$ as commuting subgroups.  
The fundamental fields $\chi^\pm_{\up , \down}$ transform in the 4-dimensional
spinor representation of $SO(5)$ and the bilinears decompose as 
${\bf 4} \otimes {\bf 4} = {\bf 1}  \oplus {\bf 5} \oplus {\bf 10}$.   
The ${\bf 5}$ vector representation serves as order parameters for the spontaneous
symmetry breaking of $SO(3)$, i.e. magnetic order,  and also contains Cooper pair
fields of charge $\pm 2$ which are order parameters for symmetry breaking of
$U(1)$, i.e. superconductivity.   By deriving separate gap equations 
and studying their solutions it was shown that the model can contain
an anti-ferromagnetic phase (AF) and a d-wave superconducting phase (SC). 
In this model the basic mechanism that leads to a d-wave SC instability 
is clearly identified as arising from the  1-loop scattering of Cooper pairs.  
An attractive feature of the model is its simplicity, reflected in the fact
that it has very few free parameters:  an overall energy scale $E_0$ set by
the high-energy cutoff $\Lambda_c$,  a Fermi velocity $v_F$,  and a single
coupling $g$.    Furthermore,  at low energies $g$ is near the low-energy
fixed point value of $1/8$,  so that perturbation theory can be reliable.    

The above features led us to propose this theory as a toy model of high temperature
superconductivity (HTSC).   Here the $SO(5)$ symmetry is quite different from
the $SO(5)$ symmetry of Zhang in this context\cite{Zhang}: whereas he postulates
an $SO(5)$ invariant Landau-Ginsburg theory for the bosonic spin and electric
order parameters,  our model is a microscopic model of fermions where the
bosonic order parameters are composite bilinears in the fermions.  
The $SO(5)$ symmetry was not put in by hand in our model, rather it is 
accidental and hidden.  
In fact,  the AF  and d-wave SC  phases are {\it not} related by
the $SO(5)$ in our model.     In principle
a Landau-Ginsburg effective theory can be derived from our model,  but like the BCS
theory,  our  
underlying microscopic theory is more powerful for deriving gap equations, 
critical temperatures, etc.    Arguments that motivate its application to HTSC
were given in \cite{KapitLeclair},  the best one being its  relation to the 
non-linear $O(3)$ sigma model effective theory for  the $2d$ Heisenberg model
description of the AF phase, which is also a relativistic theory;  
 however a rigorous derivation of it from say
the Hubbard model on the lattice  is  lacking.     If our model really turns out to 
correctly capture some  essential features of HTSC,  then it should be viewed
as an effective low-energy theory for  wavelengths that are long compared
to the lattice spacing.   One expects that such a description can be approximately
described by a rotationally invariant quantum field theory describing the 
gas of quasi-particles that SC condenses out of.   The shortcoming of such 
a theory is that lattice effects,  such as features that are dependent
on the detailed structure of the Fermi surface in the Brillouin zone,  
are necessarily absent in the basic model,  
although perhaps they can be incorporated
with small perturbations of the hamiltonian. 
Furthermore, our model does not have an explicit Fermi surface,
as for the $O(3)$ sigma model of the AF phase at half-filling.
Nevertheless, the existence of a simple model of fermions with only 
repulsive quartic interactions that possesses the all of the main motifs of
HTSC, which is to say it has clearly identified mechanisms for
the d-wave SC, the pseudogap, and anti-ferromagnetism,  can be useful
for deciding which  properties are essential and which are superfluous
for the phenomenon.   
 Thus, although it has  not  been  
established that our model captures  what is really happening in HTSC, 
for the remainder of this paper we will take the liberty to borrow the 
terminology of the HTSC literature where appropriate.

\begin{figure}[htb] 
\begin{center}
\hspace{-15mm}
\psfrag{A}{${\rm doping}$}
\psfrag{B}{$T_c$}
\psfrag{C}{$T^*$}
\includegraphics[width=10cm]{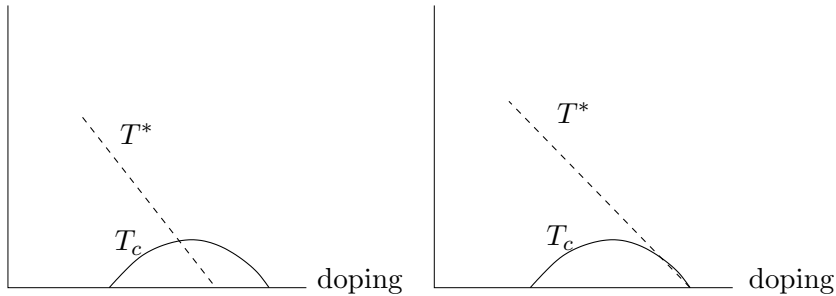} 
\end{center}
\caption{Two proposed theoretical scenarios for the HTSC phase diagram.} 
\vspace{-2mm}
\label{Figure1} 
\end{figure}

The main purpose of the present article is study in detail the origin and
properties of the so-called pseudogap in our model.   In the HTSC materials,
the pseudogap refers to an energy scale $E_{pg} \equiv k_B T^*$ where 
a cross-over behavior is observed in a variety of physical properties such
as electronic specific heat,  magnetic susceptibility and conductivity.  
The same energy scale can be observed as the onset of a depression in
the density of states.    In the underdoped region,  $T^*$ is considerably
larger than the superconducting $T_c$.  
 For reviews see \cite{PseudogapReviews1,PseudogapReviews2}.   
The origin and physical interpretation of the pseudogap remains a fundamental
question in the physics of HTSC,  and many researchers feel that a proper
understanding of it will be an important key toward unraveling the mysteries
of HTSC.    Two broad classes of theories can be summarized by the schematic
phase diagrams shown in Figure \ref{Figure1}.    In the scenario on the left, 
$T^*$ goes to zero inside the SC dome, perhaps terminating at a quantum critical
point.   In this class of theories the pseudogap is unrelated, and in fact competes with,
superconductivity.    In the second class of theories,  the pseudogap indicates 
pre-formed Cooper pairs for example, i.e. is a friendly precursor to SC,  and the pseudogap line
merges with $T_c$ on the overdoped side.    These opposing scenarios are discussed
in some detail in \cite{friendorfoe}.  
Thermodynamic data such as specific heat favors the first scenario\cite{Loram,Loram2}, 
whereas spectroscopic data seems to support the second\cite{Hufner}.
% \textbf{Eliot: Would be good to cite Kondo et al,
% Nature 457, 296 (2009) as evidence opposing preformed pairs. }
%\cite{Kondo}

In our model there is an obvious identification of the pseudogap:  whereas 
AF or SC order are related to the order parameters in the ${\bf 5}$ vector
of $SO(5)$,  the pseudogap is naturally associated with the singlet 
bilinear in the tensor product of ${\bf 4} \otimes {\bf 4}$,  as 
suggested by Tye\cite{Tye}, and also more tentatively in \cite{KapitLeclair}.     This corresponds
to the operator 
$\chi^- \chi^+ \equiv \sum_{\alpha = \up , \down} \chi^-_\alpha \chi^+_\alpha$.     
Whereas this term is classically zero in the relation with the $O(3)$ sigma model,
it is dynamically generated in the presence of interactions.   As a contribution to  the hamiltonian,
it corresponds to a relativistic mass $m$,  and the 1-particle states have energy
$E_\kvec = \sqrt{\kvec^2 + m^2}$.      Thus a non-zero mass $m$ gives a clean gap
in the density of states, i.e. there is nothing ``pseudo'' about it.  
As discussed above,  the anisotropy of the observed pseudogap is 
not a feature of this mass-gap since it is a lattice effect (``Fermi arcs''). 
In the sequel we identify this mass with the pseudogap energy
scale $E_{pg} = T^* = m$ and will show that its properties closely 
parallel the phenomenology
of the pseudogap in HTSC, at least on the underdoped side.

The pseudogap has an entirely different origin than the SC gap in our model,  essentially 
because it corresponds to  a dynamically generated vacuum expectation value
$\langle  \chi^- \chi^+ \rangle$  which preserves
the $SO(5)$ symmetry whereas SC breaks the electric $U(1)$ subgroup.
It is an intrinsic property of the normal state density of states. 
   Whether it competes with or aids superconductivity is
straightforwardly addressed by incorporating it into the d-wave gap equation 
derived  in \cite{KapitLeclair}.     As we will see,  it clearly competes with superconductivity
since if it is too large it destroys the solution to the SC gap equation.    Fortunately,
the pseudogap goes to zero at the critical point and this makes superconductivity possible;
in fact SC is enhanced near the critical point.      
Thus our model is  in the class of the left figure 
 in  Figure  \ref{Figure1},   with $T^*$ terminating
at a quantum critical point near optimal doping.   
(The proper treatment of temperature presented in this paper 
led to the identification of the critical point which differs
from the original proposal in \cite{KapitLeclair}.)

Our analysis of the pseudogap required two technical improvements of the work 
presented in \cite{KapitLeclair}.    In the latter work,  temperature was treated 
crudely as a mass,  the idea being that temperature and a mass can both be viewed
as an infra-red cutoff;    this is ultimately unreliable for the computation of 
thermodynamic properties.     
  In the present work,   since the mass is identified with 
the pseudogap which is a zero temperature property,   temperature must be dealt with
properly,  and this is accomplished with the Matsubara formalism.    A non-zero mass
also resolves some infra-red divergences that were present in the previous treatment.

The remainder of this article is organized as follows.   In the next section we review
the definition of the model and some of its basic properties. 
Section III addresses the unitarity of the model, beyond what
is contained in our previous work.  
    Section IV describes
our RG prescriptions.     In section V  we  analyze the dynamical generation of
a mass,  i.e. pseudogap,   by deriving a gap equation for the vacuum expectation value
$\langle \chi^-  \chi^+  \rangle$.       The pseudogap depends on the variable $x$ which
up to a scale is the inverse coupling.       In section VI we propose a relation between
the variable $x$ and hole doping which thus gives the doping dependence of the
pseudogap.      In section VII  we compute the effect of the pseudogap on the electronic
specific heat.      A non-zero magnetic field is introduced in section VIII,  and we compute
the temperature dependent spin susceptibility and the magnetic field dependence of
the specific heat.       For all of these thermodynamic properties,   we find crossover
behavior at the temperature $T=T^*$ and its  qualitative dependence on doping 
compares favorably with data.      Finally in section IX  we derive the finite temperature
version of the d-wave gap equation and incorporate the effect of the pseudogap into it.
Analysis of this equation clearly shows that the pseudogap competes with SC,
and  leads to a computation of the superconducting $T_c$.

\section{Review of the model,  it's symmetries and order parameters}

As for any second-quantized description of electrons with spin $\inv{2}$, 
the fundamental fields  of the model are 4 fermionic  fields $\chi^\pm_\alpha$,  where the 
flavor index $\alpha = \up, \down$ corresponds to spin and $\pm$ is electric
charge.    Due to the fermionic statistics there is a unique quartic interaction,
thus  various models are primarily characterized by the free kinetic term.  
Our model in 
two spatial dimenions  is
\beq
\label{ham}
H = \int d^2 \xvec  \( \sum_{\alpha = \up , \down}
(  \d_t \chi^-_\alpha \d_t \chi^+_\alpha 
+ v_F^2 \gradvec \chi^-_\alpha \cdot \gradvec \chi^+_\alpha + m^2 \chi^-_\alpha
\chi^+_\alpha )  + 8 \pi^2 g \,  \chi^-_\up \chi^+_\up \chi^-_\down \chi^+_\down \) 
\eeq
The above hamiltonian would be a standard second-quantized field theory for
fermions interacting via a delta-function potential if it weren't for the
term that is second order in time derivatives,  and this is the primary
novelty of the model.  
This choice of kinetic term can be motivated from the phenomenology of HTSC,
since it leads to the correct temperature dependence of the specific heat
$C \propto T^2$ at low temperatures (see sectin VII)
in the abscence of superconductivity, which is characteristic of a relativistic theory,
and the mass $m$ can correspond to the pseudogap.

As a model of HTSC,  the above kinetic term  
 can also be motivated as follows\cite{KapitLeclair}.  
Suppose one is near the Mott-Hubbard insulating phase.  
The anti-ferromagnetic 
phase of the Heisenberg model has an effective low energy description in terms of 
a spin 3-vector field $\phivec$ constrained to be of fixed length $\phivec
\cdot \phivec  = {\rm constant}$, 
with lagrangian
 $\d_\mu \phivec \cdot \d_\mu \phivec$\cite{Haldane,Affleck}.    
The field $\phivec$ is bilinear in the fundamental electron fields, 
 and in our model
corresponds to $\phivec = \chi^-  \sigmavec \chi^+ / \sqrt{2}$.   
The constraint on $\phivec$ can be imposed by the constraint $\chi^- \chi^+ = {\rm constant}$
since 
\beq
\label{1.1b}
\phivec \cdot \phivec  = - \frac{3}{2} \( \chi^- \chi^+ \)^2
\eeq
Imposing these constraints one finds that 
\beq
\label{1.1c}
\d_\mu \phivec \cdot \d_\mu  \phivec ~ \propto ~  \d_\mu  \chi^- 
 \d_\mu \chi^+   ~~~ + 
{\rm irrelevant~  operators}
\eeq
which justifies the kinetic term in our model.     One can then relax the constraint
on $\chi^- \chi^+ $, 
  and replacing  it with a ``soft constraint'' by including a
quartic interaction,  as is done for the 
non-linear $O(3)$ sigma model in two spatial dimensions.   

Another motivation for the kinetic term in our model (at $m=0$) 
was given in \cite{KapitLeclair}
based on the linear dispersion relation $E_\kvec = |\kvec|$ one obtains when expanding around a circular
Fermi surface.  However this involved a modification of, or at best a crude approximation to,
 the density of states.      
  
  Canonical quantization and also a path-integral formulation
follow from   
       the  euclidean action:
\beq
\label{1.1}
S = \int d^2 \xvec \, dt   \(  \sum_{\alpha = \up, \down} \( \d_\mu  \chi^-_\alpha \d_\mu \chi^+_\alpha  
+ m^2  \chi^-_\alpha \chi^+_\alpha \)   -  8 \pi^2  g  \,    \chi^-_\up \chi^+_\up  \chi^-_\down \chi^+_\down  \)    
\eeq
where $\d_\mu \d_\mu =  \d_t^2  + v_F^2  \gradvec^2$.   The velocity $v_F$ plays the role
of the speed of light, and it was proposed in \cite{KapitLeclair} that it be identified
with the universal nodal Fermi velocity\cite{Nodal}.   In the sequel we set $v_F = \hbar = k_B = 1$
except where indicated.   
Note that the fields $\chi$ are treated as Lorentz scalars and spin is
simply a flavor.   However it is possible to treat the fields as Dirac
spinors and thereby achieve complete Lorentz invariance\cite{Dean}. 
The quartic term  is unique up to the sign of the coupling by fermionic statistics,
and positive $g$ corresponds to repulsive interactions.

A consequence of the fermionic statistics is that the model has a hidden $SO(5)$ symmetry. 
As explained in the Introduction, 
the appearance of this $SO(5)$ is quite different from the $SO(5)$ symmetry proposed
by Zhang.  
This symmetry is manifest if one considers an $N$-component version with fields
$\chi^\pm_\alpha$,  $\alpha = 1,.., N$,
which has $Sp(2N)$ symmetry (hence the terminology ``symplectic fermions'').
 For $N=2$, since there are 4 fermionic fields
and consequently a unique 4-fermion interaction,  the theory automatically
has $Sp(4) = SO(5)$ symmetry.   

The $SO(5)$ contains
   $SO(3)$  and  $U(1)$ subgroups which commute and can be identified with 
spin and electric charge respectively.     The conserved  electric current 
then corresponds to 
\beq
\label{current}
J^e_\mu = -i   \sum_\alpha \(  \chi^-_\alpha  \d_\mu  \chi^+_\alpha  +  \chi^+_\alpha \d_\mu  \chi^-_\alpha  \)  
\eeq
and the fields $\chi^\pm$ have electric charge $Q_e= \pm 1$.

The important order parameters for the study of spontaneous symmetry breaking are composite  bilinears 
in the fermions.    The 4 fields $\chi^\pm_\alpha$ transform under the spinor representation 
of $SO(5)$.    The bilinears can be decomposed as ${\bf 4} \otimes {\bf 4} =  {\bf 1}  \oplus 
{\bf 5}   \oplus {\bf 10}$ where  $ {\bf 1}$  is the singlet,  ${\bf 5}$ the vector representation, 
and ${\bf 10}$ the adjoint.      The singlet is the field   $\sum_\alpha \chi^-_\alpha  \chi^+_\alpha  \equiv  \chi^-  \chi^+  $
and corresponds to the mass term in the action.    The ${\bf 5}$-vector  of fields 
corresponds to 
\beq
\label{1.2}
\vec{\Phi}  =   (\phivec ,   \phi^+_e  ,  \phi^-_e  )    =  (\inv{\sqrt{2}} \chi^- \sigmavec \chi^+   ,  
\chi^+_\up \chi^+_\down ,  \chi^-_\down \chi^-_\up  )
\eeq
where $\sigmavec$ are Pauli matrices.    The triplet of fields $\phivec$ are electrically neutral
and transform as a spin vector under the $SO(3)$ and serve as magnetic order parameters.  
The fields $\phi^\pm_e$ on the other hand are spin singlets but carry electric charge $\pm 2$
and are thus Cooper pair fields for superconducting order.     
The $SO(5)$ invariant product is
\beq
\label{1.3} 
\vec{\Phi} \cdot \vec{\Phi}  =   \phivec \cdot \phivec -  2  \phi^+_e  \phi^-_e  
\eeq
and the interaction can be expressed in the manifestly $SO(5)$ invariant manner:
\beq
\label{1.4}
\CL_{\rm int}   =   \frac{8 \pi^2 }{5}  g  \,   \vec{\Phi} \cdot \vec{\Phi}  
\eeq

The momentum expansion of the free fields is
\barray
\label{momexp}
\chi^-   (\xvec , t)  &=&   \int   \frac{d^2 \kvec}{(2\pi)^2 \sqrt{2 \omega_\kvec} } 
\(  a^\dagger_\kvec  e^{-i k \cdot x}   +   b_\kvec  e^{i k \cdot x}    \) 
\\ \nonumber
\chi^+  (\xvec , t)  &=&  \int   \frac{d^2 \kvec}{(2\pi)^2 \sqrt{2 \omega_\kvec}}  
\( - b^\dagger_\kvec  e^{-i k \cdot x}   +   a_\kvec  e^{i k \cdot x}    \) 
\earray
where $\omega_\kvec =  \sqrt{\kvec^2 + m^2}$ and $k\cdot x =  \omega_\kvec t - \kvec \cdot \xvec$.    
The canonical quantization of the theory based on the lagrangian
 leads to the canonical anti-commutations in
momentum space:
\beq
\label{aadag}
\{ a_\kvec ,  a^\dag_{\kvec'}  \}   =   \{  b_\kvec ,  b^\dag_{\kvec'} \}  =  (2 \pi)^2 \delta (\kvec - \kvec' ) 
\eeq
and the hamiltonian is 
\beq
\label{hamilton}
H_{\rm free} 
 =  \int   \frac{d^2 \kvec}{(2\pi)^2 }   \sum_{\alpha = \up , \down}  \omega_\kvec  \,   
\(  a^\dag_{\kvec, \alpha} a_{\kvec, \alpha}   +  b^\dag_{\kvec, \alpha}  b_{\kvec , \alpha}  \)   
\eeq
The $a$ and $b$ particles have opposite electric charge and can thus
be thought of  as particles and holes.

A distinguishing feature of our model, in contrast to quartic interactions
of Dirac fermions for instance, is that the quartic interaction is relevant
in the renormalization group (RG) sense:   the field $\chi$ has classical 
mass dimension $1/2$ so that the quartic interaction has dimension $2$
and the coupling $g$ dimension 1.    This means that the interactions
are important at low energies  and can lead  to non-Fermi liquid
behavior.   In fact,   the model has a low energy RG fixed point, i.e.
a quantum critical point at $g\approx 1/8$.  This critical point can
formally be understood as an analytic continuation of 
 the familiar Wilson-Fisher fixed point of
the $O(N)$ models,  where $N=-4$,  and some critical exponents, such as
the anomalous dimension of the field $\chi$,  can be computed  by specializing 
 the known epsilon-expansion results for the  $O(N)$ model  to
$N=-4$\cite{Neubert}.      However, many of the important operators such as the order parameters $\Phivec$ are composite fermion bilinears, which changes their structure compared to the magnetic order parameters of the $O(N)$ models. Furthermore,
 $\Phivec$ includes bilinear order parameters of charge $\pm 2$ 
that have no counterpart in $O(N)$ physics.
These RG properties are summarized in the next section. 

Another interesting feature of our  model is that whereas the fundamental
interaction is repulsive,   when one incorporates the momentum-dependent
scattering at second order in perturbation theory (1-loop)  there is
an instability toward the formation of a d-wave superconducting ground state.
This was studied in \cite{KapitLeclair} by deriving a d-wave gap equation
for momentum-dependent vacuum expectation values for the Cooper-pair fields
$\phi^\pm_e$.   This is reviewed briefly in section IX where we derive
and study the finite temperature version of the d-wave gap equation.

\section{Unitarity of the model as an effective low energy theory.}

The above free hamiltonian (\ref{hamilton}) in momentum space 
is obviously hermitian and defines a unitary theory,
in spite of the fact that the  Klein-Gordon type of action
 normally associated with 
free bosons was quantized with the ``wrong'' statistics, i.e. as a fermion. 
There are no negative norm states in the Fock space of $a, b$ particles
since there are no unwanted minus signs in eq. (\ref{aadag}). 
In $1d$  the free massless symplectic fermion model 
 is normally considered a non-unitary $c=-2$ conformal field theory.
The resolution of this apparent contradiction can be found in \cite{Guru}, 
where it was shown that the symplectic fermion 
can be mapped onto the $c=1$  unitary theory\cite{Guru} of Dirac fermions
at the level of detailed conformal partition functions.  
In light of the above results,  the latter fact is not surprising
since in momentum space the hamiltonians 
of symplectic and Dirac fermions are identically eq. (\ref{hamilton}).   
   
The issue of unitarity was further addressed  in \cite{Neubert, KapitLeclair}, where  
the  main concern is the consistency of  the interacting theory.
Because of the extra minus sign in (\ref{momexp}), the fields
$\chi^\pm$ are not hermitian conjugates of each other,  but rather
\beq
\label{unit.1}
\chi^+  =  C (\chi^-)^\dagger C 
\eeq
where the unitary operator $C$ simply distinguishes particles and holes:
$C a C = a$ and $C b C = -b$,  with $C^\dagger C = C^2 = 1$. 
The operator $C$ is easily constructed:
\beq
\label{Ccons}
C = \exp \(  i \pi \int \frac{d^{2} \kvec }{(2\pi)^2} \sum_\alpha b^\dagger_{\kvec, \alpha}
b_{\kvec , \alpha} \)
\eeq
The interacting theory
is thus pseudo-hermitian:
$H^\dagger = C H C$.  
The free
theory is actually hermitian in momentum space since it is quadratic in $b$'s. 
As before there are no negative norm states in the Fock space.

A pseudo-hermitian hamiltonian has real energy eigenvalues,
and can still lead to a unitary time evolution,
as discussed in  \cite{Bender,Mostafazadeh,Neubert,KapitLeclair}. 
 Let us make the following
additional remarks concerning this issue in this specific context. 
   A unitary time evolution
follows if  one modifies the definition of hermitian conjugation.
Define the $C$-hermitian conjugate  as $A^\daggerc = C A^\dagger C$.
Then the S-matrix is unitary with respect to this conjugation:
$S^\daggerc S = 1$ since $H^\daggerc = H$.  
  However  the $C$-conjugate hermitian conjugation
implies  a modified definition of the inner product: 
$\langle \psi' | \psi \rangle \to 
\langle \psi' | \psi \rangle_c \equiv  \langle \psi' | C |\psi \rangle$.  
Under this new inner product,  states with an odd number of $b$ particles
now have negative norm.  A completely 
equivalent description that dispenses altogether with  the operator $C$ 
amounts to removing the 
minus sign in the expansion of $\chi^+$ in eq. (\ref{momexp}).
This gives a hermitian hamiltonian since now 
$\chi^- = (\chi^+)^\dagger$.  However this also 
gives  the modified relation
$\{ b_\kvec , b^\dagger_{\kvec'} \} = - (2\pi)^2\delta (\kvec - \kvec')$ 
and this leads to the same conclusion that the $b$ particle sector contains
negative norm states.  
  
We now argue that this model can still yield  a consistent  theory
at 
 {\it low energy}.   Let us use the description based on the $C$-operator;
the same arguments apply to the manifestly hermitian description without it. 
The probability that an initial  state $|i\rangle$ evolves to a final 
state $| f \rangle$
is given by 
\beq
\label{prob}
P_{if}  =  \frac{ |\langle i | S |f \rangle_c |^2}{\langle i | i \rangle_c 
\langle f | f \rangle_c}
\eeq
States can be classified according to their $C$ eigenvalue of $\pm 1$,
where $C=-1$ corresponds to negative norm.   One sees from the above
formula that the probability $P_{if}$ is positive so long as 
the states $|i\rangle , | f \rangle $ are both of either positive
or negative norm.   The problematic transitions with negative probability
thus arise if the matrix element $\langle  i | S |f \rangle$ is 
non-zero for states of mixed norm.  
However at low energies compared to the mass  $m$, these transitions 
are forbidden  as a result of energy and charge conservation.  For example,
  suppose the initial
state  consists of only $a$-particles.  Because of charge conservation,
processes that change the number of $b$ particles necessarily involve
creation of $a,b$ pairs;  for instance $aa \to ab$ is not allowed.
The process $a \to a a b$ is consistent with charge conservation,
but  not allowed kinematically,
so one needs to consider at least $a a \to a a a b$.   Energy conservation
shows that the energy  of the initial $a$ particles must be at least 
$2m$ in order for the process to be kinematically allowed.   
 Thus at  low energies compared
to the mass $m$  no $b$-particles will be
produced from a state containing only $a$ particles. 
    A variation of this viewpoint interchanges 
the role of $b$ and $b^\dagger$, so that particles are created by
the $b$'s and have negative energy.  One can now imagine that 
all the negative energy states are filled according to the Pauli
principle.  Since there is a gap of $2m$ between the negative and
positive energy states,  at low energies compared to $m$ no 
$ab$ pairs will be created out of the Dirac sea and one can deal
only with the $a$-particles.   

As we describe in the sequel,  the mass $m$ is on the order of 
the cut-off and can be quite large, typically greater than
$1000K$ for HTSC materials, so the low energy approximation is expected to 
be good for a large range of temperatures.   At very high temperatures,
our model necessarily breaks down and must be replaced by another
effective theory that restores unitarity. 
For instance, the cross-over to ordinary Fermi liquid behavior
in for example the specific heat is not built into the model. 

Another way to obtain a low energy unitary theory involves
the  introduction of a chemical potential,  which is necessary
at finite density.   Particle number is not conserved here; instead
a chemical potential couples to the electric charge $Q_e$, i.e. 
$H \to H + \mu Q_e$.   
The free hamiltonian is then
\beq
\label{Hmu}
H = \int \frac{d^2 \kvec}{(2\pi)^2} \(  (\omega_\kvec + \mu ) a_\kvec^\dagger
a_\kvec + (\omega_\kvec - \mu ) b^\dagger_\kvec b_\kvec \) 
\eeq
Because of the difference in electric charge between the $a$ and $b$ particles,
at a given $\kvec$ there is a gap of $2\mu$ between the energies of
the $a$'s and $b$'s.    Let us suppose that $\mu$ is chosen so that
all of the $b$-states are filled.  Typically $\mu$ will be 
comparable to the Fermi energy and typically very large.  If $\mu$ is large enough, 
then  because of the energy gap
$2\mu$ at low energies we can effectively deal only with the $a$-particle
sector.    Alternatively,  one can view the $a$ sector as an 
in-accessible high energy sector when $\mu$ is large enough, 
and deal only with the low-energy $b$ sector.   Either way,
as discussed above, all probabilities $P_{if}$ will be positive.

\section{Renormalization group prescriptions}

In all approaches to the renormalization group    there are two cut-offs, 
 a fixed upper cut-off $\Lambda_c$
typically  related to the inverse lattice spacing, 
  and a lower cut-off $\Lambda < \Lambda_c$ 
which is the running RG scale.     Renormalization effectively  removes the degrees of freedom with
energy between $\Lambda$ and $\Lambda_c$.   In the high-energy physics context,   the cut-off
$\Lambda_c$    is unknown,   so one normally performs the renormalization by sending $\Lambda_c$ to
infinity and incorporating counter-terms to cancel ultra-violet divergences.    In $D=4$   spacetime dimensions
this can be conveniently and systematically performed by using dimensional regularization in $D= 4 - \epsilon$
dimensions,   which leads to the epsilon-expansion technique for studying models in $D=3$.   
In the present context,  since the cut-off $\Lambda_c$ is in principle known and physical quantities
depend on it,  it is more appropriate to perform the RG directly in $D=3$ with explicit cut-offs rather
than use the  epsilon expansion.   

 The dependence of the couplings on the running RG scale $\Lambda$ 
can be determined  by considering an upper cut-off $\Lambda$ only.   Consider first the
coupling constant $g$.    The 1-loop correction to the vertex is cancelled by a shift of $g$, 
i.e. by defining $g(\Lambda)   =  g + \delta g$ where
\beq
\label{2.2}
\delta g  =   16 \pi^2  g^2  \int^\Lambda   \frac{d^3 \ell}{ (2\pi)^3}   \inv{ (\ell^2 + m^2 )^2 }  
\eeq
When $\Lambda \gg m$,   $g(\Lambda) =  g - 8 g^2 / \Lambda$.    Thus
\beq
\label{2.3}
\Lambda  \d_\Lambda g(\Lambda)  =  \frac{8 g^2}{\Lambda}   
\eeq
Let us define 
\beq
\label{2.5}
g  =  \Lambda  \ghat 
\eeq
Then the beta function for the dimensionless coupling $\ghat$ is 
\beq
\label{2.6} 
\Lambda \d_\Lambda \ghat =  - \ghat +  8  \ghat^2  
\eeq

The above beta-function eq. (\ref{2.6}) has a low energy fixed point at $\ghat = 1/8$.
It is known that this fixed point survives, with small corrections, 
 to higher orders\cite{Neubert}.     
We can now integrate the RG flow and incorporate the initial data at the high energy
cut-off $\Lambda_c$.      
It will be  convenient   to introduce the dimensionless variable $x$:
\beq
\label{2.7}
x \equiv  1/\ghat 
\eeq
The fixed point occurs at $x_* = 8$.   
Let $\ghat_0$ be the value of the coupling at the scale $\Lambda_c$,
and $x_0 = 1/\ghat_0$.    
%The dependence on the initial data $\ghat_0$ can be characterized by
%the parameter $\gamma = (x_* - x_0)/x_*$,   where $0 < \gamma < 1$.    
Then the solution to the RG flow equation (\ref{2.6}) takes the simple 
linear form in $x$:
\beq
\label{2.8}
\frac{\Lambda}{\Lambda_c}   =   \frac{x_* - x}{x_* - x_0}  , ~~~~~
(x_* = 8 )
\eeq

There are two cases to consider depending on whether the coupling
is strong $(x_0 < x_*)$ or weak $(x_0> x_*)$ at short distances.   
Based on the relation between $x$ and hole doping described in
section VI,  we will refer to $x_0 < x < x_*$ as the underdoped
region and $x_* < x < x_0$  as overdoped;  in both cases 
$\Lambda > 0$.      Furthermore,  
in order to clearly distinguish the two cases we will refer to
$x_0$ in the overdoped region as $\tilde{x}_0$.  
(What we refer to as the overdoped region is close but not identical
to the usual terminology;  the latter refers to the region
beyond the maximum $T_c$.)

It is clear that our model is characterized by a single fundamental energy
scale set by the cutoff $\Lambda_c$.    Since $\Lambda_c$ has units of a 
wave-vector $\kvec$, this energy scale  is
\beq
\label{energyscale}
E_0 = \hbar v_F  \Lambda_c  \equiv  k_B T_0
\eeq
where $\Lambda_c$ should be proportional to the inverse lattice spacing.

Before turning to mass renormalization,  
  we now determine the 1-loop corrections to
the anomalous dimension of the order parameters which are composite fields. 
    Consider first the singlet operator $\chi^- \chi^+$.   
 To  first order in perturbation theory,   
 \beq
 \label{2.23}
 \langle  \chi^- \chi^+  (0)  \cdots \rangle =  8 \pi^2 g   
 \[   \int  \frac{d^3 \ell}{(2\pi)^3}  \inv{ (\ell^2 + m^2)^2 }  \] 
 \langle  \chi^-  \chi^+ (0) \cdots  \rangle   
 \eeq
 This contribution is cancelled  by   renormalizing the operator as follows:
 \beq
 \label{2.24}
 \chi^- \chi^+   \to   Z_{\chi^- \chi^+}  \,   \singlet 
 \eeq
 where $Z_{\singlet}  =  1 + \delta Z_{\singlet}$ and
 \beq
 \label{2.26}
 \delta Z_{\singlet}   =   -  8  \pi^2 g  \int^\Lambda   
    \frac{d^3 \ell}{(2\pi)^3}  \inv{ (\ell^2 + m^2)^2 } 
 \eeq
 The anomalous dimension 
 $\gamma_{\singlet}$  of $\singlet$ is then 
 \beq
 \label{2.27}
 \gamma_{\singlet}  =  \Lambda \d_\Lambda  \log Z_{\singlet}   =  - 4 \ghat 
 \eeq
 where again  we have assumed $\Lambda \gg m$.   
Repeating the above calculation for the $SO(5)$ vector of order parameters one finds
\beq
\gamma_{\vec{\Phi}}  =   4 \ghat 
\eeq

Let $\dim{\CO}$ denote the scaling dimension of $\CO$ at the fixed  point,  including the
classical contribution of $1/2$ for each $\chi$ field.     Using $\ghat _* = 1/8$ at the fixed point,
one obtains to 1-loop
\beq
\label{2.29}
\dim{\chi^-  \chi^+}   \approx   \inv{2} ,   ~~~~ \dim{\Phivec} = 
\dim{\chi^-  \sigmavec \chi^+ }  =   \dim{\chi^+_\up \chi^+_\down}  =  \dim{\chi^-_\down \chi^-_\up}  \approx   \frac{3}{2}
\eeq
Thus one sees that the singlet is shifted down from the classical value,  whereas the 5-vector of order
parameters is shifted up.      The above values agree exactly with the 1-loop approximation 
in the epsilon-expansion obtained in \cite{Neubert}.   We point out that although the actual values
of  the fixed point value  $g_*$ and the functions  $\gamma (g)$  differ in the calculation performed here
compared to the epsilon expansion,  the results for the anomalous dimensions
at the critical point  agree to 1-loop.   
From the two-loop results obtained in \cite{Neubert},  one sees that higher order corrections
are reasonably  small:   $\dim{\singlet}$ changes to $5/8$  and $\dim{\vec{\Phi}}$  actually remains at
$3/2$ at the next order.

\begin{figure}[htb] 
\begin{center}
\hspace{-15mm}
\includegraphics[width=6cm]{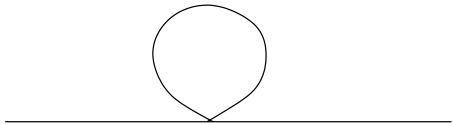} 
\end{center}
\caption{One-loop correction to the self-energy.} 
\vspace{-2mm}
\label{Figure2} 
\end{figure}

Now we consider mass renormalization,  which will be important in the sequel.    The 1-loop correction 
to the self-energy shown in Figure \ref{Figure2} is a negative correction to  $m^2$.      This is cancelled 
by $m^2 \to m^2 (\Lambda) =  m^2 + \delta m^2 $ where
\beq
\label{2.20}
\delta m^2  =  8 \pi^2  g  \int^\Lambda   \frac{ d^3 \ell}{ (2 \pi)^3 }   \inv{ \ell^2 + m^2 }     
\eeq
The $m^2$ term on the RHS is extracted by taking a derivative with respect to $m^2$ and then
setting $m^2 = 0$:
\beq
\label{2.21}
\delta m^2 = 4 g m^2 / \Lambda
\eeq
Defining  
\beq
\label{mhat}
m = \Lambda  \mhat 
\eeq
one finds
\beq
\Lambda \d_\Lambda  \mhat^2  =   (-2  -  4 \ghat )  \mhat^2 
\eeq
The $-2 \mhat^2 $ term simply corresponds to classical dimension $2$ of $m^2$
and $4 \ghat$   the quantum correction.    Since  $m^2$ is the coupling for
$\chi^-  \chi^+$ this implies that the anomalous dimension of $\singlet$ is  $-4 \ghat$,
in agreement with eq. (\ref{2.27}).

\section{Dynamical mass generation and its identification with the pseudogap}

Although the mass $m$ is classically zero in the 
connection to the $O(3)$ sigma model description
of the Heisenberg AF described in section II,
the operator $\singlet$  is relevant and  not forbidden by any symmetries so that  in general
it will be generated when one includes interactions.   
In this section we determine this dynamically generated mass to lowest order 
and propose that it be identified with the HTSC pseudogap.  

  If the original theory is massless, 
at  1-loop the correction to $m^2$ coming from the diagram in Figure \ref{Figure2}
 is 
negative  and equal to $- 8 \pi^2 g \int \frac{d^3 \ell}{ (2 \pi)^2 } \inv{\ell^2}$.    Since the
propagator $1/\ell^2$ becomes $1/(\ell^2 + m^2)$ when $m\neq 0$,   a self-consistent
equation for $m$ is the following
\beq
\label{3.4}
m^2  =  - 8 \pi^2 g \int   \frac{d^3 \ell}{  (2 \pi)^3 }  \inv{\ell^2 + m^2 }   
\eeq
The one point function for the singlet is then
\beq
\label{3.5} 
\langle \singlet \rangle =  -2  \int   \frac{d^3 \ell}{  (2 \pi)^3 }  \inv{  \ell^2 + m^2 }   
=  \frac{m^2}{4 \pi^2 g} 
\eeq
where we have used the equation (\ref{3.4}).   The two above equations can 
also be derived from a Hubbard-Stratanovich transformation based on the fact that
the interaction is proportional to $(\singlet)^2$,  as was done in \cite{Tye}.    

In order to make sense of eq. (\ref{3.4}) and obtain solutions,   the mass renormalization 
discussed in section IV needs to be taken into account.    Taking the limits of integration to
be from zero to $\Lambda$,  eq.  (\ref{3.4}) becomes
\beq
\label{3.6}
m^2  =  -4 g \Lambda + 4 g m \tan^{-1} \Lambda / m 
\eeq
From eq. (\ref{2.20}) with $m=0$ on the right hand side, 
one obtains $\delta m^2 = 4 g \Lambda$,   therefore
the first term on the RHS above can be absorbed into $m^2$,  and this is consistent
with renormalization.      The equation for $m$ now becomes 
\beq
\label{3.7}
\mhat =  4 \ghat \tan^{-1} 1/\mhat 
\eeq
and has real solutions which are easily found numerically.    

There are two useful analytic limits to the solution of eq. (\ref{3.7}).   When $\ghat$ is large,
$\mhat$ is also large and the solution is approximately:
\beq
\label{3.8}
\mhat \approx  2 \sqrt{ \ghat} ,  ~~~~~ \langle \singlet \rangle \approx \frac{\Lambda}{\pi^2}   ~~~~~~~~(\ghat ~{\rm large})
\eeq
When $\ghat$ is small,  $\mhat$ is also small and $\tan^{-1}1/\mhat \approx \pi/2$.   Thus in this limit
one has
\beq
\label{3.9}
\mhat \approx  2 \pi \ghat,  ~~~~~\langle \singlet \rangle \approx g  ~~~~~~~~~(\ghat ~ {\rm small})
\eeq
It is interesting to note that if one sends the cut-off to infinity in  eq. (\ref{3.4}) and performs the integral
by analytic continuation in the spacetime dimension, then one obtains the solution (\ref{3.9}).   
This means that the analog of dimensional regularization here overestimates $\mhat$,  since
by eq. (\ref{3.7}),  $\mhat <  2 \pi \ghat$.     In the underdoped region the approximation
(\ref{3.8}) is considerably better than (\ref{3.9}) and we will use it in places below.

The mass $m$ corresponds to an $x$-dependent energy scale 
\beq
\label{Epg}
E_{pg} (x)  = \hbar v_F  m(x)  \equiv  k_B T^* (x) 
\eeq
where as before $x=1/\ghat$ and $\mhat (x)$ is the solution to eq. 
(\ref{3.7}). 
In units of the fundamental scale $E_0$:
\beq
\label{Epg2}
\frac{E_{pg}}{E_0}  =  \frac{T^*}{T_0}  =   \mhat (x)  \frac{\Lambda}{\Lambda_c} 
= \mhat(x) 
\frac{\xstar -x}{\xstar -x_0} 
\eeq
In the next section we will relate  $x$ to doping,  thus the energy scale
$T^*$ is doping dependent.  It should be emphasized that $T^*$ simply corresponds
to the temperature independent energy scale $E_{pg}$ and is thus not a real temperature;
however as we  will show in the sequel it can correspond to a cross-over scale 
in the real temperature.

A non-zero mass $m$ clearly corresponds to a gap in the density of
states since the 1-particle energies are $E_\kvec = \sqrt{\kvec^2 + m^2}$. 
We discuss this further at the end of this section.   
As a  model of HTSC we thus  identify the mass $m$  with
the pseudogap energy scale, i.e.   $m=E_{pg} = T^*$.
As we will show below,  the thermodynamic properties
of our model also support the identification of $T^*$ with the HTSC pseudogap.

\begin{figure}[htb] 
\begin{center}
\hspace{-15mm}
\psfrag{Y}{$T^*/T_0$}
\includegraphics[width=10cm]{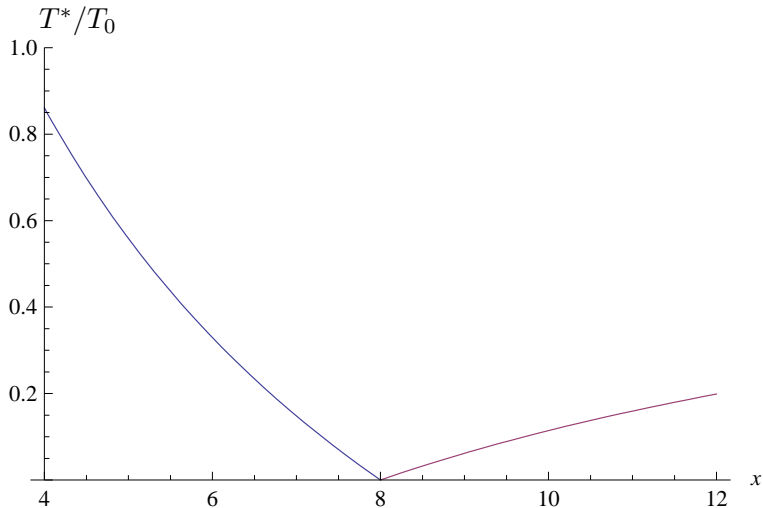} 
\end{center}
\caption{The pseudogap $T^*$ as a function of $x$ for $x_0 = 4$ and $\tilde{x}_0 = 16$.} 
\vspace{-2mm}
\label{Figure3} 
\end{figure}

A plot of $T^*$ verses 
$x$ is shown in Figure \ref{Figure3}.  
 Note that $T^*$ is close to linear near the critical
point at $x_* = 8$.   Furthermore, 
the pseudogap is smaller on the overdoped side. 
We point out that the re-appearance of the pseudogap on the overly
doped side is contrary to what is normally observed.

Since  the overall energy scale $E_0$ depends both on $v_F$ and $\Lambda_c$,
in attempting to compare with the HTSC data it is more useful to fix 
$E_0$ by using the pseudogap.  It is known experimentally that at zero
doping in the region of the AF phase $E_{pg}$ is approximately the
AF exchange energy $J$.  Since zero doping occurs at $x_0=0$ (see the next
section),  we can identify $E_{pg} (x_0) = \mhat (x_0) E_0 \approx J$.
Since $\mhat (x_0)$ is of order 1,  $E_0 \sim J$.  
For the cuprates $J/ k_B \sim  1300-1400K$.

\begin{figure}[htb] 
\begin{center}
\hspace{-15mm}
\psfrag{A}{$\rho(E)$ }
\psfrag{B}{$\frac{k_F}{2\pi}$}
\psfrag{C}{$T^*$}
\psfrag{D}{$E$} 
\includegraphics[width=10cm]{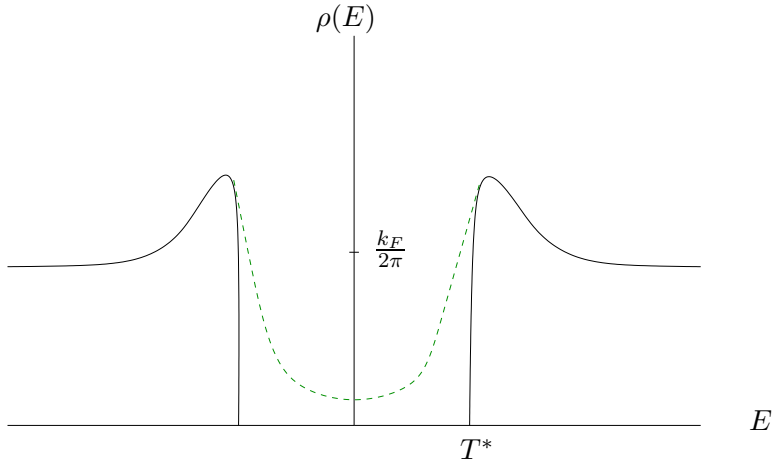} 
\end{center}
\caption{Density of states as a function of energy $E$. The dotted line
represents  a sketch of what is measured using ARPES.} 
\vspace{-2mm}
\label{Figure4} 
\end{figure}

An estimate of the density of states $\rho (E)$ as a function of energy $E$
can be obtained as follows.   One has
\beq
\label{den.1}
\rho (k) dk =  \frac{d^2 \kvec}{(2\pi)^2} =  \frac{k dk}{2\pi} = \rho(E) dE 
\eeq
If we approximate $k$ in the above equation as the Fermi wave-vector 
$k_F$, then 
\beq
\label{den.2}
\rho (E) \approx \frac{k_F}{2\pi} \( \frac{d E}{dk} \)^{-1}  
=  \frac{k_F}{2\pi}  \inv{\sqrt{1- m^2/E^2}}
\eeq
This behavior is qualitatively  shown in Figure \ref{Figure4}.

Angle-resolved photoemission (ARPES) measurements show a clear depression
in the density of states and the leading edge of this depression
is characterized by the pseudogap scale $T^*$.  This behavior is
shown schematically in Figure \ref{Figure4} as a dotted line.  Thus one qualitative
difference with what is observed is that the gap in our model is a clean one.
Another difference is that due to the rotational invariance of our model,
the pseudogap is isotropic,  whereas the observed pseudogap shows lattice
effects.   It is anisotropic
in the direction relative to the Fermi surface:  the pseudogap is largest in
the anti-nodal $(0, \pi)$ direction and vanishes along ``Fermi arcs'' in 
the nodal $(\pi , \pi)$ direction.

\section{Hole doping}

Since renormalization removes degrees of freedom between $\Lambda$ and $\Lambda_c$,
 this suggests 
that doping can be varied by varying the cutoff $\Lambda$.   In \cite{KapitLeclair}  it was argued that a measure
of hole doping can be defined based on the 1-point function $\langle \singlet \rangle$.   Since
this one point function is proportional to $\Lambda$ by dimensional analysis,  we are led to identify doping $p$ with 
the dimensionless quantity 
\beq
\label{hole.1}
p  =   \frac{ c }{\Lambda_c}  \(  \langle \singlet \rangle_{\Lambda_c}  -   \langle \singlet
 \rangle_\Lambda \)  
\eeq
for some constant $c$.    In the underdoped regime, the 1-point function is approximately given 
by eq. (\ref{3.8}):
\beq
\label{hole.2}
p(x) \approx 
  \frac{c}{\pi^2}  \(  1 -  \frac{\Lambda}{\Lambda_c} \)  =
  \frac{c}{\pi^2}   \( 
\frac{x-x_0}{x_* - x_0}  \)  
\eeq
where we have used eq. (\ref{2.8}).    
Half-filling then corresponds to $\Lambda = \Lambda_c$,  i.e. $x=x_0$.     
Thus, in the approximation we have made,  plots of various physical properties as a function of doping
$p$ is simply related to plots as a function of the inverse coupling
 $x$ by rescaling and shift of the $x$-axis.    We 
choose to plot  against $x$ since  this  more clearly reveals the RG properties.

We can give a rough estimate of the constant $c$ following an argument made in 
\cite{KapitLeclair}.    Consider a lattice fermion model where $\vec{S}_\xvec$ is the
local spin variable at lattice site $\xvec$,  which is bilinear in the fermion
operators.  One has
\beq
\label{hole.3}  
\vec{S} \cdot \vec{S} = - \frac{3}{2} n_\up n_\down + \frac{3}{4} ( n_\up + n_\down ) 
\eeq
where $n_{\up, \down}$ is the number of fermions of spin up or down at each site.   
From this relation one sees that at half-filling  $n_\up n_\down =0$, 
$n_\up + n_\down =1$ and $\vec{S}^2 = 3/4$,  i.e. $\vec{S}$ is constrained to be
a spin $\inv{2}$ vector and the model can be mapped onto the Heisenberg model.   
Dividing the above equation by the volume squared and taking the infinite volume
limit, one finds that the right hand side is $-\frac{3}{2} \rho_\up \rho_\down$
where $\rho_{\up, \down}$ are number densities.   

In our continuum model,  $\vec{S}$ is represented by the bilinear $\phivec$
and one has the identity:
\beq
\label{hole.4}
\phivec \cdot \phivec  = - \frac{3}{2} (\chi^- \chi^+ )^2 
\eeq
Comparing with the continuum limit of eq. (\ref{hole.3}),  one identifies
$\chi^- \chi^+ =  (\rho_\up + \rho_\down)/\sqrt{2}$, which corresponds to 
$c = \sqrt{2}$.    
At the critical point this gives $p_{\rm crit} \approx .14$. 
  Since this is only a rough estimate,  one can alternatively
fix $c$ in principle by fitting to experimental data.

\section{Specific heat}

An approximation to the specific heat can be made based on including just the 
effects of the dynamically generated mass.
      A gas of 4 types of fermionic particles 
with single particle energies $\omega_\kvec = \sqrt{\kvec^2 + m^2 }$  has the
free energy per volume at temperature $T$:
\beq
\label{sp.1}
\CF = -4 T \int_0^{\Lambda_c} 
  \frac{d^2 \kvec}{(2\pi)^2} \log \( 1 +  e^{-\omega_\kvec /T} \)
\eeq
One easily sees that $\CF / \Lambda_c^3$ is a function of
the two dimensionless variables $ T/ T_0 $ and $T/T^*$ where as in 
section V we have identified $T^* = m$.     

At very low temperatures $T\ll T_0$ 
we can approximate the $\log$ as $e^{-\omega_\kvec /T}$
and effectively send the cut-off $\Lambda_c$ to infinity,
and $\CF/T^3$ becomes  a scaling function of $T^*/T$.   
The result is 
\beq
\label{sp.2}
\CF \approx   - \frac{2 T^3}{\pi}   e^{-T^*/T} \( 1 + \frac{T^*}{T} \)   
\eeq
The entropy density is then:
\beq
\label{sp.3}
S = - \frac{\d \CF}{\d T}  \approx  \frac{3 T^2}{\pi} e^{-T^*/T} \( 1 + 
\frac{T^*}{T} + \inv{3}
\( \frac{{T^*}}{T} \)^3 \)  
\eeq
and the specific heat:
\beq
\label{sp.4}  
C =  - T \frac{\d^2 \CF}{\d T^2}  \approx \frac{12 T^2}{\pi}  e^{-T^*/T} \(
1 + \frac{T^*}{T}  + \inv{2} \(\frac{{T^*}}{T}\)^2  + 
\inv{2} \( \frac{{T^*}}{T}\)^3  \) 
\eeq
It is convenient to define the quantity $\gamma = C/T$ since for
a Fermi liquid $\gamma$ is a constant.   The primary feature of 
our model is that $\gamma \propto  T$ at very low temperatures compared
to $T^*$ and $T_0$,  which is ultimately attributed to the 
relativistic nature of the model.

At higher temperatures there is a crossover to a different 
 behavior.     Since $\Lambda_c$
can be scaled out of the eq. (\ref{sp.1}),   the cross-over temperature 
is the pseudogap temperature $T^*$.     Below we plot 
the entropy and $\gamma$ as a function $T$ for various ``doping''  $x$.   
The plots are in terms of the dimensionless quantities $T/T_0$,
$S/\Lambda_c^2$ and $\gamma/\Lambda_c$.  
One clearly sees the crossover at $T=T^*$.

\begin{figure}[htb] 
\begin{center}
\hspace{-15mm}
\psfrag{C}{$S$}
\psfrag{A}{$T/T_0$}
\includegraphics[width=10cm]{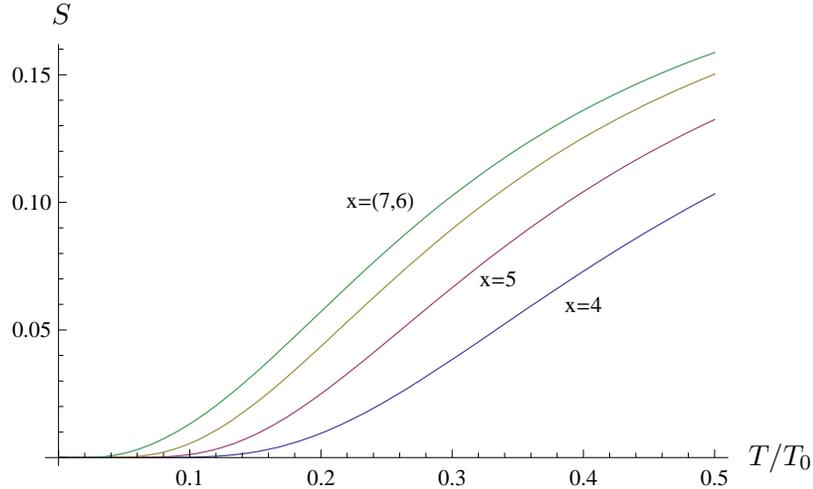} 
\end{center}
\caption{Entropy as a function of temperature for $x_0 = 4$.  
The vertical axis is the dimensionless quantity 
$S/E_0^2$.} 
\vspace{-2mm}
\label{Figure5} 
\end{figure} 

\begin{figure}[htb] 
\begin{center}
\hspace{-15mm}
\psfrag{A}{$T/T_0$}
\psfrag{D}{$\gamma$}
\includegraphics[width=10cm]{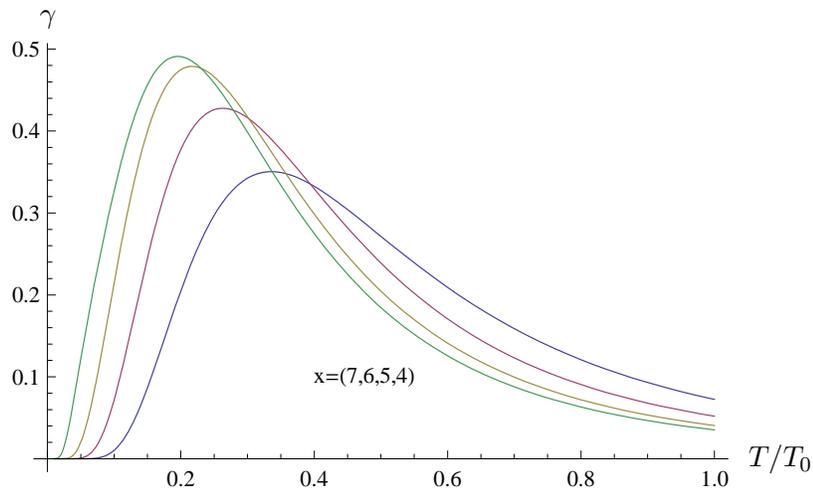} 
\end{center}
\caption{$\gamma = C/T$ as a function of temperature at various
``doping'' $x$ for $x_0 = 4$.  The larger amplitude curves correspond
to larger doping. 
The vertical axis is the dimensionless quantity $\gamma / E_0$.} 
\vspace{-2mm}
\label{Figure6} 
\end{figure}

\begin{figure}[htb] 
\begin{center}
\hspace{-15mm}
\includegraphics[width=10cm]{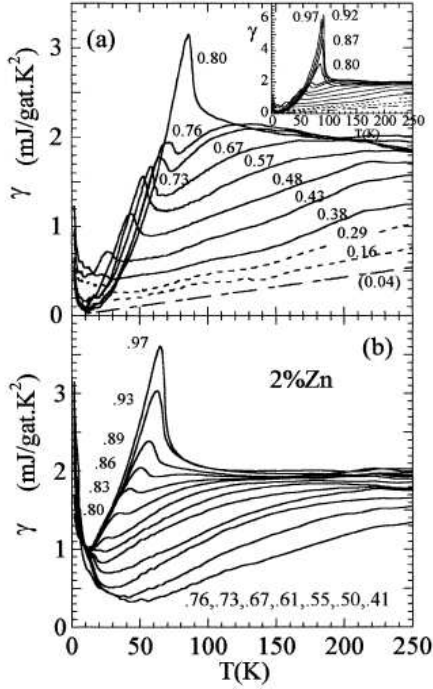} 
\end{center}
\caption{Experimental data for $\gamma (T)$ for 
${\rm Y_{0.8}Ca_{0.2} Ba_2 Cu_3 O_{6+x}}$ (from \cite{Loram}).} 
\vspace{-2mm}
\label{LoramFig1} 
\end{figure}

Experimental data for $\gamma$ is shown in Figure \ref{LoramFig1}.
In comparing with our results, it is important to bear in mind
that the data contains contributions from the quasi-particle
excitations in the SC phase,  which explains the peaks 
to the left,  and such effects are not included in 
our calculation.   Thus,  our curves should be compared with
the data to the right of the SC peaks.     
 At temperatures
below $T_0$ one sees a reasonably good qualitative agreement with
the behavior computed above:  for $T\ll T^*$,  $\gamma \propto T$, 
with a crossover to a different behavior at $T^*$.  The dependence on
doping is also qualitatively correct,  i.e. the peaks of the curves
move toward the left with increased doping.
 One difference is that whereas our computed $\gamma$ goes
to zero at high temperature,  the experiments indicate that at high
temperatures $\gamma$ approaches the Fermi-liquid result,  i.e. 
$\gamma$ approaches a constant.   Our model of course does not crossover
to a Fermi liquid at high temperatures, because as explained in section 
III it is expected to break down at high enough temperatures, 
  so at best it may describe
 temperatures up to and above $T^*$.

\section{Non-zero magnetic field}

In this section we compute the spin response to a magnetic field
and also the magnetic field dependence of the specific heat in the
same approximation we made in the last section, i.e. we only consider
the effects of the  dynamically generated mass.  
Both these quantities can be studied by adding a term to the
action $\int d^2 \xvec dt  ~ \sqrt{2} \vec{h} \cdot \phivec$.  
Comparing with the calculation of the effective potential 
$V_{\rm eff}  = \CF$ in \cite{KapitLeclair}, one sees that  
the free energy at zero temperature is  
\beq
\label{mag.1}
\CF =  -  \int \frac{d\omega d^2 \kvec}{(2\pi)^3}  ~
\log \(  (\omega^2 + \omega_\kvec^2)^2 - h^2 \)
\eeq
where as before, $\omega_\kvec^2 = \kvec^2 + m^2$.

\subsection{Spin response}

The one point function $\langle \phivec \rangle$ in the
presence of $\vec{h}$ can be obtained from the first derivative
of the logarithm of the partition function $Z= \exp( - \CF V/T )$
where $V$ is the volume.    For a single component of $\phivec$, 
one obtains 
\beq
\label{mag.2} 
\langle \phi \rangle = \inv{\sqrt{2}}  \frac{\d \CF}{\d h} 
= \inv{\sqrt{2}} \int \frac{d\omega d^2\kvec}{(2\pi)^3} 
\( \inv{\omega^2 + \omega_\kvec^2 - h}  -  \inv{\omega^2 + \omega_\kvec^2 + h}
\)
\eeq

At finite temperature $T$, $\omega$ becomes a quantized Matsubara
frequency $\omega = 2\pi \nu T$ where $\nu \in Z + 1/2$ and one makes
the replacement:
\beq
\label{mag.3}
\int \frac{d\omega}{2\pi}  \to   T \sum_{\nu \in Z + 1/2}  
\eeq
We need the identity
\beq
\label{mag.4}
T \sum_{\nu \in Z + 1/2} \inv{a^2 + (2\pi \nu T)^2 }  =   \inv{2a} 
\tanh \( \frac{a}{2T} \) 
\eeq
This gives
\beq
\label{mag.5}
\langle \phi \rangle = \inv{2\sqrt{2}} \int \frac{d^2 \kvec}{(2\pi)^2 } 
\(  \inv{\sqrt{\omega_\kvec^2 -h }}  \tanh  \frac{\sqrt{\omega_\kvec^2 -h}}
{2T}    - ( h \to -h) \)
\eeq

\begin{figure}[htb] 
\begin{center}
\hspace{-15mm}
\psfrag{B}{$\chi_s$}
\psfrag{A}{$T/T_0$}
\includegraphics[width=10cm]{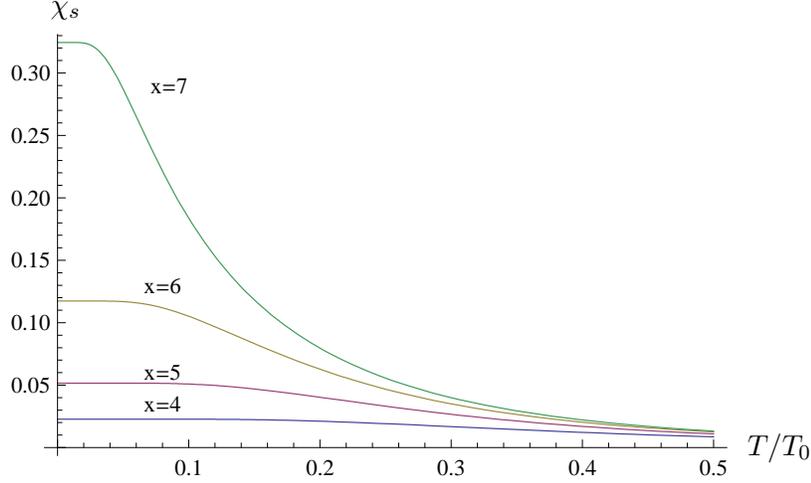} 
\end{center}
\caption{The spin response as a function of temperature at various
doping variable $x$  for $x_0 = 4$.
The vertical axis corresponds to the dimensionless quantity 
$\chi_s  \cdot E_0$.} 
\vspace{-2mm}
\label{Figure7} 
\end{figure}

 The linear response for small $h$  follows from Taylor expanding the 
integrand:
\beq
\label{mag.6} 
\langle \phivec \rangle =  \chi_s (m, T) \, \vec{h}
\eeq
where
\beq
\label{mag.7}
\chi_s (m,T) =  \inv{4\sqrt{2} T}  \int \frac{d^2 \kvec}{(2\pi)^2} 
\inv{\omega_\kvec^2}  \(  \tanh^2 \frac{\omega_\kvec}{2T} 
+ \frac{2T}{\omega_\kvec} \tanh \frac{\omega_\kvec}{2T} - 1 \)
\eeq
Plots of $\chi_s$ as a function of temperature are shown in Figure \ref{Figure7}.  
One sees that the spin response is quenched when $T< T^*$,  consistent
with experiments.

\subsection{Specific heat}

Define 
\beq
\label{mag.8}
\Delta \CF =  \CF(h) - \CF (0)
\eeq
Integrating the identity eq. (\ref{mag.4}) one can show
\beq
\label{mag.9}
T \sum_{\nu \in Z + 1/2} \(  \log (\omega_\nu^2 + \omega_\kvec^2 -h ) 
- \log (\omega_\nu^2 + \omega_\kvec^2) \)  = 
2 T \log \( \frac{1+ 
e^{-\sqrt{\omega_\kvec^2 -h}/T}} {1 + e^{-\omega_\kvec /T}} \)
+ \sqrt{\omega_\kvec^2 - h} - \omega_\kvec 
\eeq
Therefore up to a constant that is independent of $T$ and does
not  affect the specific heat:
\beq
\label{mag.10}
\Delta \CF = -2T \int  \frac{d^2 \kvec}{(2\pi)^2}  
\log \( \frac{ (1+ e^{-\sqrt{\omega_\kvec^2 - h}/T})(1 + 
e^{-\sqrt{\omega_\kvec^2 + h}/T} )} {(1 + e^{-\omega_\kvec /T})^2 }
\) 
\eeq
One can then define
\beq
\label{mag.11}
\Delta \gamma = -  \frac{ \d^2 \Delta \CF}{\d T^2}
\eeq

\begin{figure}[htb] 
\begin{center}
\hspace{-15mm}
\psfrag{gdiffy}{$\Delta \gamma \left( T \right)$}
\psfrag{gdiffx}{$T/T_{0}$}
\includegraphics[width=10cm]{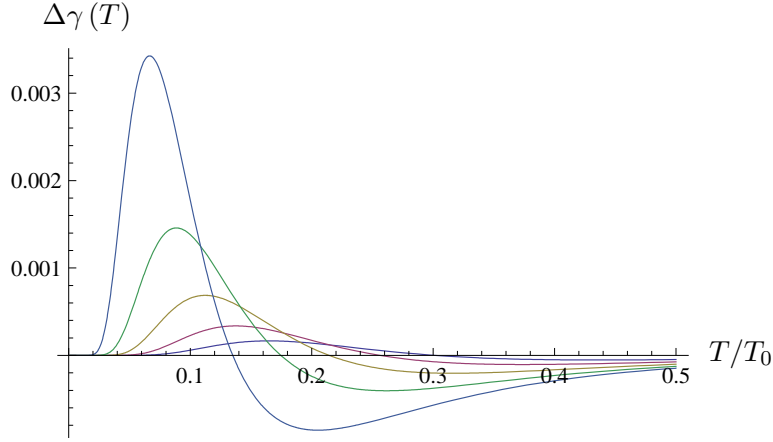} 
\end{center}
\caption{$\Delta \gamma (h,T)$ evaluated at 
$x=4, 4.5, 5, 5.5, 6$ for $x_0=4$
 as a function of temperature for fixed $h=0.025$.
The larger amplitude curves correspond to larger $x$, i.e. 
larger doping.} 
\vspace{-2mm}
\label{Figure8} 
\end{figure}

\begin{figure}[htb] 
\begin{center}
\hspace{-15mm}
\psfrag{gdiffy}{$\Delta \gamma \left( T \right)$}
\psfrag{gdiffx}{$T/T_{0}$}
\includegraphics[width=10cm]{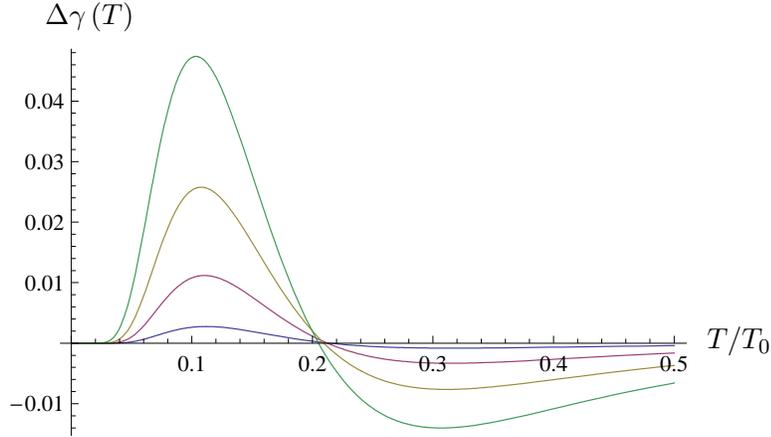} 
\end{center}
\caption{$\Delta \gamma (h,T)$ as a function of temperature for various $h = 0.05,0.1 ,0.15,0.2$ (increasing $h$ corresponds to 
increasing amplitudes on the plot) at fixed $\left( x, x_{0} \right) = \left( 5,4 \right)$.} 
\vspace{-2mm}
\label{Figure8b} 
\end{figure}

Plots of $\Delta \gamma$ as a function of $T$ for  different $x,h$ 
are shown in Figures \ref{Figure8}, \ref{Figure8b}. 
(These plots are in arbitrary units since we have not specified
the strength of the magnetic field.)   
At very small $h$,  $\Delta \gamma \propto h^2$.

%\begin{figure}[htb] 
%\begin{center}
%\hspace{-15mm}
%\includegraphics[width=10cm]{LoramgammaH.eps} 
%\end{center}
%\caption{Experimental data on $\Delta \gamma (h,T)$ as a function of temperature 
%for various $h = ?, ? , ?$ at fixed doping (from \cite{LoramMag}).} 
%\vspace{-2mm}
%\label{deltagamExp} 
%\end{figure} 

\section{Finite temperature d-wave gap equation and the effect of the pseudogap.}    

\subsection{Zero temperature}

It was shown in \cite{KapitLeclair}  that the 1-loop corrections to scattering
of Cooper pairs leads to a d-wave superconducting gap 
of the form
\beq
\label{d.0}
q(\kvec ) = \delta^2 ( k_x^2 - k_y^2 ) = \delta^2 k^2 \cos (2\theta)  
\eeq
where $q(\kvec)$ is a Fourier transform of Cooper pairing order parameters
$\langle \chi^\pm _\up (\xvec_1 ) \chi^\pm_\down (\xvec_2 )\rangle$.  
Before turning to the main topic,  
some remarks on the d-wave property of the gap are called for.   
Since the d-wave gap equation derived in \cite{KapitLeclair} was
based on a rotationally invariant hamiltonian,   the most general
solution involves a rotation by an arbitrary angle $\theta_0$, i.e.
$\cos 2 \theta   \to \cos 2 (\theta - \theta_0)$ in the above equation.  
On the other hand in HTSC the lattice breaks the rotational symmetry
and the d-wave gap is oriented with respect to the Fermi surface
in a specific way,  namely the gap vanishes in the nodal direction
$(k_x , k_y) = (\pi , \pi)$,  which corresponds to $\theta_0 = 0$.   
This can be reconciled as follows.   Free particles on the 
lattice have dispersion relation $\varepsilon_\kvec = -2 (
\cos k_x  + \cos k_y )$.  Thus the Fermi surface has the symmetry
$k_x \to -k_x$.   Requiring this symmetry fixes $\theta_0 = 0$.

In this sub-section we study the effect of the non-zero mass $m$  on the
zero temperature  d-wave
superconducting gap equation.   
The effect of the mass term is simply the shift 
$\omega^2 \to \omega^2 + m^2$ and the d-wave gap equation derived
in \cite{KapitLeclair} becomes:
\beq
\label{d.1}
\delta^4 = 2 g_2 \int d\omega dk^2 \( 
1-  \frac{\omega^2 + k^2 + m^2}{\sqrt{ (\omega^2 + k^2 + m^2 )^2 
+ \delta^4 k^4 }}  \)
\eeq
where $k^2 = \kvec^2 $.   
The coupling $g_2$ comes from the 1-loop scattering of Cooper pairs
and is given by
\beq
\label{d.2}
g_2  =  \frac{8 \pi^2 g^2}{5} \int  \frac{d^3 \ell }{(2\pi)^3}  
\inv{ (\ell^2 + m^2 )^4 }
\eeq
For simplicity we incorporate the cut-off as follows:
$|\omega| < \infty $ and $k^2 < \Lambda_c^2$ which 
 is more appropriate for comparison
with the finite temperature version we consider below.

Re-expressing the gap equation in terms of dimensionless 
quantities by rescaling  
 $k \to \Lambda_c k$ and $\omega \to \Lambda_c \omega$, one obtains 
\beq
\label{d.3}
\delta^4 = 2 \ghat_2 \int_0^\infty   d\omega \int_0^1 dk^2 \( 
1-  \frac{\omega^2 + k^2 + m'^2}{\sqrt{ (\omega^2 + k^2 + m'^2 )^2 
+ \delta^4 k^4 }}  \)
\eeq
where $m' = m/\Lambda_c$ and $\ghat_2  =  g_2 \Lambda_c^3$.   

\begin{figure}[htb] 
\begin{center}
\hspace{-15mm}
\psfrag{ylabel}{$\delta' \left( x \right)$}
\psfrag{xlabel}{x}
\includegraphics[width=10cm]{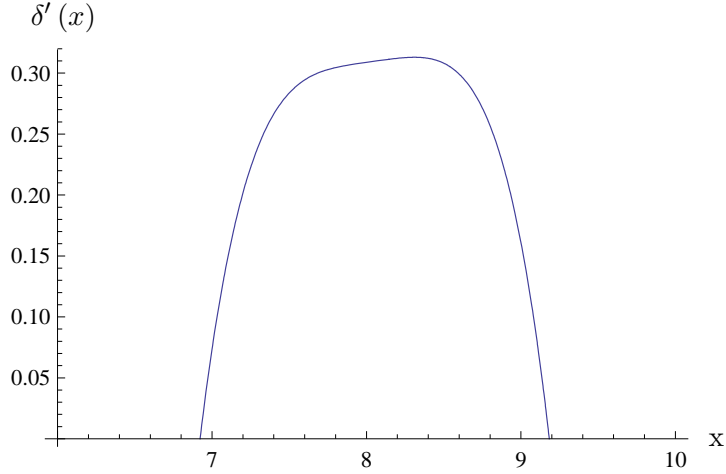} 
\end{center}
\caption{Plot of $\delta_{q}' \left( x \right)$ vs $x$ for $x_{0} = 4$ and 
$\tilde{x}_{0} = 12$.} 
\vspace{-2mm}
\label{deltax} 
\end{figure}

Since eq. (\ref{d.2}) is ultra-violet convergent,  it can be
approximated by letting the upper cut-off go to infinity, 
giving $g_2 = \pi g^2 / 40 m^5$.    Incorporating the RG prescriptions
of section IV,   $g= \Lambda \ghat$ and $m = \Lambda \mhat$, 
one finds that the parameters in the gap equation (\ref{d.3}) 
are the following:
\beq
\label{d.4}
\ghat_2  =  \frac{\pi}{40} \frac{\ghat^2}{\mhat^5} \( 
\frac{\Lambda_c}{\Lambda} \)^3 , ~~~~~ m' =  \frac{\Lambda}{\Lambda_c} \mhat
\eeq
where the ratio $\Lambda/ \Lambda_c$ is given in eq. (\ref{2.8}). 
    Finally,  since under a RG transformation $\delta k \to  \delta \Lambda
k / \Lambda_c  \equiv \delta' k$,   the physical gap in the theory
at RG scale $\Lambda$ is  $\delta' = \Lambda \delta / \Lambda_c$.
The $x$-dependence of the solutions $\delta$ arises from 
the $x$-dependence of $\ghat_2$ and $m'$.  
A plot of $\delta'$ as a function of x  is shown in Figure \ref{deltax}.
It is important to point out that the interpretation of the 
critical point at $x_*$ presented here differs from the original
proposal in \cite{KapitLeclair}, in that in the present work 
we extend $x$ beyond $x_*$ by introducing $\tilde{x}_0$, 
and this places the maximum value of the gap near the critical point
$x_*=8$.  (We have effectively patched together what was referred to as 
Type A and B in \cite{KapitLeclair}.)  
Further justification for this location of the critical 
point is based on the calculation of $T_c$, which reaches 
a maximum value near the critical point, as explained below.

One can easily verify numerically that the pseudogap competes with 
SC in the following sense:   if one artificially increases the mass $m$,
then the value of the gap $\delta$ decreases,  and eventually vanishes
for $m$ too large.  (This was shown already in \cite{KapitLeclair}.)

\subsection{Finite temperature}

In order to derive the finite temperature version of the above 
d-wave gap equation,  we start with the un-integrated form 
derived in \cite{KapitLeclair}:
\beq
\label{ft.1}
q(\kvec)  =  -  \int  \frac{d\omega d^2 \kvec'}{(2\pi^3)} 
G(\kvec , \kvec' )  \frac{ q(\kvec')}{ (\omega^2 + \omega_{\kvec'}^2 )^2 
+ q(\kvec')^2 }
\eeq
where the kernel is
\beq
\label{ft.2}
G(\kvec, \kvec') = - 8 \pi^2 g_2 k^2 k'^2 \cos 2(\theta - \theta')
\eeq
The equation (\ref{d.1}) is obtained upon performing the angular 
integral.  

At finite temperature,  $\omega$ becomes a quantized Matsubara
frequency $\omega_\nu = 2 \pi T \nu$, where $T$ is the temperature
and $\nu$ is a half-integer, i.e.  $\nu \in Z + 1/2$.  
As before, the integral $\int d\omega /2\pi$ is replaced with $T\sum_\nu$.   
One needs the identity:
\beq
\label{ft.3}
T \sum_\nu  \inv{ (\omega_\nu^2 + \omega_\kvec^2)^2 + q^2 } =
 \frac{T}{q}  \Im \sum_\nu \inv{\omega_\nu^2 + \omega_\kvec^2 - i q }
= \inv{q}  \Im \(  \inv{2 \omega_{\kvec,q}} \tanh \( 
\frac{\omega_{\kvec, q}}{2T} \)  \) 
\eeq
where $\omega_{\kvec, q} = \sqrt{\omega_\kvec^2 - i q }$.
   
Due to the specific form of the kernel $G$, the solution 
to the equation (\ref{ft.1})  is of the d-wave form (\ref{d.0}) (up to
an arbitrary rotation)  
where $\delta$ satisfies the integral equation
\beq
\label{ft.4}
\delta^2 =  g_2  \int dk d\theta ~ k^3 \cos(2\theta) 
\Im \(  \inv{\omega_{\kvec,\delta}} \tanh \( 
\frac{\omega_{\kvec, \delta }}{2T} \)  \) 
\eeq
where 
\beq
\label{ft.5}
\omega_{\kvec , \delta}  =  \sqrt{ \omega_\kvec^2 - i \delta^2 k^2 
\cos 2\theta }
\eeq
Finally,  the physical temperature at the scale $\Lambda$ follows
from the RG transformation $T \to T \Lambda / \Lambda_c$.   

\begin{figure}[htb] 
\begin{center}
\hspace{-15mm}
\psfrag{ylabel}{$\delta' (T)$}
\psfrag{xlabel}{$T/T_0$}
\includegraphics[width=10cm]{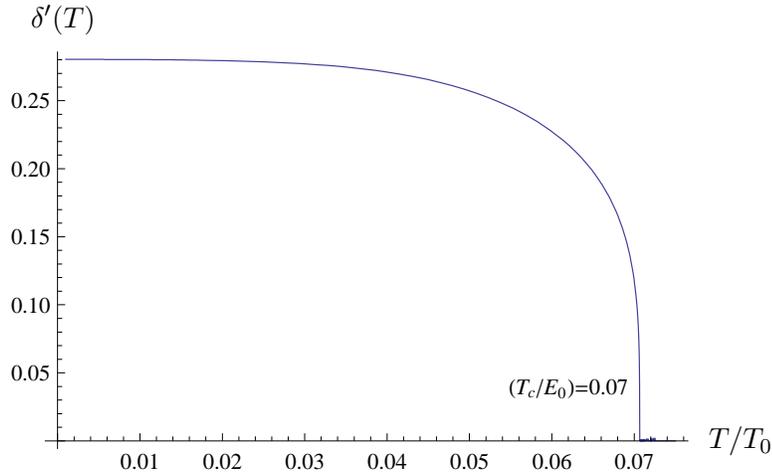} 
\end{center}
\caption{The d-wave gap $\delta' (T)$ as a function of $T$ for $x=7.5,
x_0 = 4$.} 
\vspace{-2mm}
\label{deltaofT} 
\end{figure}

Solutions of the above equation are $\delta (x,T)$.   One can easily verify
numerically 
that as the temperature goes to zero, one recovers the solution $\delta (x)$ 
to the zero temperature gap equation (\ref{d.3}).  One also finds that
as the temperature is raised there are no solutions to the above equation
for $T>T_c$ and this defines the $x$-dependent 
critical temperature $T_c$.   This is 
shown in Figure \ref{deltaofT}.

\begin{figure}[htb] 
\begin{center}
\hspace{-15mm}
\psfrag{ylabel}{$T_{*}/T_0,T_{c}/T_0$}
\psfrag{xlabel}{x}
\includegraphics[width=10cm]{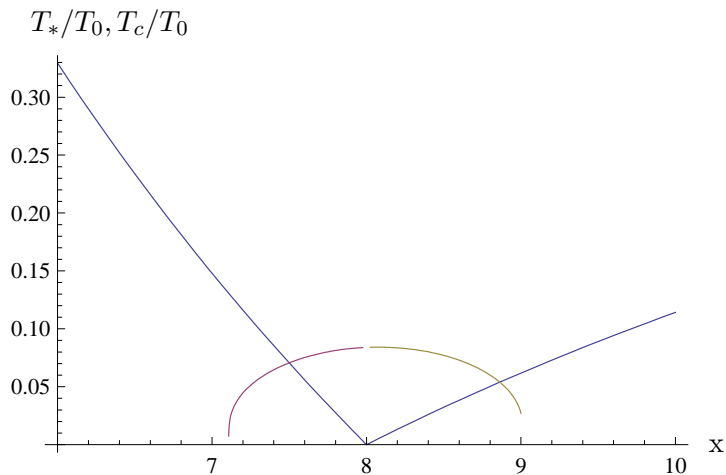} 
\end{center}
\caption{Plot of $T_{c}$ and  $T_{*}$ vs $x$ 
for $x_{0} = 4$ and $\tilde{x}_{0} = 12$.} 
\vspace{-2mm}
\label{Tc} 
\end{figure}

The critical temperature $T_c$ as a function of $x$ on both sides
of $x_*$ are shown in Figure \ref{Tc}.  It turns out that 
$T_c$  at the critical value $x=x_*$, i.e. $T_c^* = T_c (x_*)$,  is universal in that 
it only depends on the overall scale $T_0$ and not on 
$x_0 , \tilde{x}_0$ since the scale factor $\Lambda / \Lambda_c$ 
vanishes at this point. 
Using the estimated relation between
$x$ and doping in section VI,  the critical point occurs at 
doping $p_{\rm crit} \approx .14$. 
The dome shape of $T_c$ is a property of the mathematical structure
of the gap equation and nothing universal is happening at
the termination points of SC on either side of the critical point. 
Numerically we find that  
at the critical point 
\beq
\label{ratio}
\frac{T_c^*}{T_0} \approx .084, ~~~~~\frac{T_c^*}{T_0 \delta' (x_*)}  \approx  .268
\eeq   
We also find numerically
that the maximum value of $T_c$ occurs close to 
the critical point so that $T_c^{\rm max} \approx .084 T_0$,
i.e.  $T_c^{\rm max}$ is simply proportional to fundamental energy
scale $E_0$.   

As argued in section V,  $T_0$ should be identified with 
the anti-ferromagnetic exchange energy $J$ at half-filling.  
For $T_0 = 1350K$,  this gives $T_c^{\rm max} \approx 113K$, 
which is quite reasonable for HTSC.      
It should be emphasized  that the above $T_c$ is  intrinsic
to the two spatial dimensions, i.e. does not involve
any kind of inter-planar energy scales. 

%\begin{figure}[htb] 
%\begin{center}
%\hspace{-15mm}
%\psfrag{Y}{$T_{c}/T_{0}$}
%\psfrag{X}{x}
%\includegraphics[width=10cm]{dwaveplot.eps} 
%\end{center}
%\caption{$T_{c}$ as a function of $x$ 
%for various $x_{0}$ and $\tilde{x}_{0}$.} 
%\vspace{-2mm}
%\label{Tc} 
%\end{figure} 

\section{Concluding remarks}

In summary,  we have further developed the interacting
symplectic fermion model in two spatial dimensions by 
studying a dynamically generated relativistic mass and
by including a finite temperature.   This allowed us to 
study some fundamental properties of the model, such 
as the specific heat and spin response,  which clearly
show non-Fermi liquid properties.    As a simplified model
of HTSC,  we identified the pseudogap energy scale with the zero temperature
relativistic mass $m$,  and pointed out some close parallels with
the observed phenomenology of the pseudogap.  

We studied the effects of the pseudogap and finite temperature
on the d-wave gap equation.  In this model, the pseudogap 
clearly competes with superconductivity as a distinct  phenomenon.
Our analysis of $T_c$ suggests that the quantum critical point of our
model, where the pseudogap vanishes,  occurs inside the superconducting dome
near optimal doping.   For an antiferromagnetic exchange energy of
$J\sim 1350K$, solutions of the d-wave gap equation give a
maximum $T_c$ of about $110K$.

\section{Acknowledgments}

We would like to thank Ian Affleck, Denis Bernard, Dean Robinson,  Henry Tye,
and Jan Zaanen  
 for
discussions.    This work is supported by the National Science Foundation
under grant number  NSF-PHY-0757868. 
EK  would also like to acknowledge the support of the National Defense Science and Engineering Graduate Fellowship of the American Society for Engineering Education.

\end{document}